# Controlling Home Appliances Remotely through Voice Command

Faisal Baig
Electrical Department
Federal Urdu University of Arts,
Science and Technology
Islamabad, Pakistan

Saira Beg
Computer Science Department
COMSATS Institute of
Information Technology
Islamabad, Pakistan

Muhammad Fahad Khan
Computer Science Department
Federal Urdu University of Arts,
Science and Technology
Islamabad, Pakistan

## ABSTRACT
Controlling appliances is a main part of automation. The main object of Home automation is to provide a wireless communication link of home appliances to the remote user. The main objective of this work is to make such a system which controls the home appliances remotely. This paper discusses two methods of controlling home appliances one is via voice to text SMS and other is to use the mobile as a remote control, this system will provide a benefit to the elderly and disable people and also to those who are unaware of typing an SMS.

## 1. INTRODUCTION
Home automation is not a new concept in today's world, it is used to provide convenience for user to remotely control and monitor the appliances and it provides a better use of electricity. The efficient use of electricity makes the HOME automation to play an important role in daily life. As by the growth of PC (personal computers), internet, mobile phone and wireless technology makes it easy for a user to remotely access and controls the appliances. A lot of research has been done and many solutions have been proposed to remotely access the HOME appliances. Some of them used internet, wireless technology to communicate and control home appliances, others used Bluetooth and GSM technology for controlling the home appliances.

Proposed method reduces the wiring and complexity of the system. It has no geographical limitation and can be used on any GSM network; it provides portability to the system. It is mainly focused on the elderly people, disables and for the people who are unable to type text or face difficulties in typing. For the disable people, it is quite difficult to operate the HOME appliances physically or they are unable or feel uncomfortable to type a text so as to switch on/off the relative device as in [1] [7] [9]. So a system has been developed to monitor the Appliances remotely by simply running the mobile application and giving voice command. The mobile application efficiently converts the voice command to text and transfers it to the GSM network. It is affordable to everyone, cheap and easy to install. As there is no wired communication between the remote user and appliances control module and the electronic devices used to control are easily available making it a cost effective solution.

The technology used to develop the system is Java for mobile and MPLAB for microchip family of controller, and Bluetooth interface for wireless communication between home mobile and hardware control module.

## 2. RELATED WORK
[1] is about controlling home appliances through a microcomputer, author discusses two different approaches to control the home appliances; approaches are timer option and voice command. The timer option provides control based on timer, and the voice command provide control by using voice commands to control the appliances. This system uses a PC and PC parallel port to control the appliances, and the software interface is developed on the VB 6.0. This is used to convert voice command in to text and provide the operation to control and monitor the appliances.

[2] Proposed a system that control home appliances through infra-red remote controller and power line communication by developing a home based server, this system help user to check the status of their appliances form anywhere through the cellular network and internet.

[3] Proposed a GSM based system for controlling the Appliances for the people who are not at home, this is done remotely through SMS over GSM network using AT commands and on receiver the GSM modem is interfaced with the PC, the home appliances control system is developed on the PC to monitor and control. In the proposed solution they use PC parallel port which is further interfaced with the rely circuit to provide control over the appliances. This system also provides a feedback by simply SMS to user which also helps when there is any security breech in the home.

[4] Proposed a solution of home appliances control using Bluetooth based remote control to access the control of home appliances within home, author developed a remote control with a Keypad which is interface to a microcontroller and this is interfaced to Bluetooth module to provide wireless interface for the remote to communicate with the appliances control module. When the key is pressed the controller send the command regarding the pressed key via Bluetooth medium and on the receiver end receiver receive the command and apply the corresponding action.

[5] The Author Proposed a system that uses a PC based application to convert voice command to txt and transmit this command via a user mobile to the cellular network, on the

receiver the mobile receive the SMS which is read by the microcontroller using AT command structure, the communication medium between the microcontroller and mobile is RS232 standard which is a wired communication. After complete reception of the command the controller perform the action.

[7] the Author present a system in which the client system is programmed with an application which is used to control and monitor the appliances, the application that is developed for the client system is convert the given voice command to the symbolic data that is to be transferred via WI-FI network to the server computer. The server computer contains the Speech recognition application developed in Microsoft Visual Basic.net. so as the communication establish between the client system and the server it start the speech recognition and when the given data is received this is transfer to the control circuit via PC parallel port and the given load is switched on.

## 3. PROPOSED METHOD

The aim of this research is to propose a model for home appliances control, the system is shown in Figure 1.

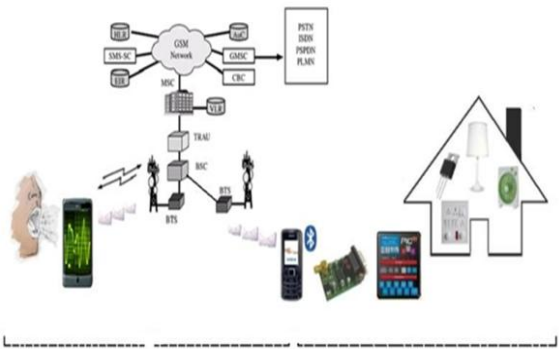

**Fig 1: proposed system Architecture**

The diagram in fig.1 shows a model from where user with an android OS based Mobile giving voice command, the mobile application convert the voice into text using android intent API 2.01. Conversion of voice into text SMS step undergoes many other steps such as;

- After taking voice command, convert it into ByteOutputStream.
- Than convert ByteOutputStream into unsigned integer array. And add 255 into all those values which lie between 0-31 ranges.

The commands generated are appended in SMS payload which is sent through GSM network. Table 1 shows the basic commands which then translate into SMS commands.

**Table 1. Voice Command and its related SMS Command**

| Voice Command to Text Command Attributes | |
|---|---|
| **Voice Commands** | **SMS Commands** |
| Main Switch On | ' SON1E' |
| Main Switch OFF | 'SOFF1E ' |
| Light On | ' LON1E ' |
| Light Off | ' LOFF1E ' |
| Fan On | ' FON1E ' |
| Fan Off | ' FOFF1E' |

### 3.1 Hardware Design

On the receiver end it consist of a mobile which is programmed with an application software, which is used to transfer the receive SMS via a Bluetooth channel to the controller of microchip family PIC16F877A. It is a 8bit processor, the Controller receive the SMS by communicating with the Bluetooth module for the reception of the SMS, the communication between the PIC16F877A and Bluetooth module is of RS232 standard. Fig. 2 shows the PIC16F877A development kit. After the complete reception of the command the controller applies the given action by sending pulse to the switching circuit, after this the microcontroller checks the status of the appliances and send a feedback to the user.

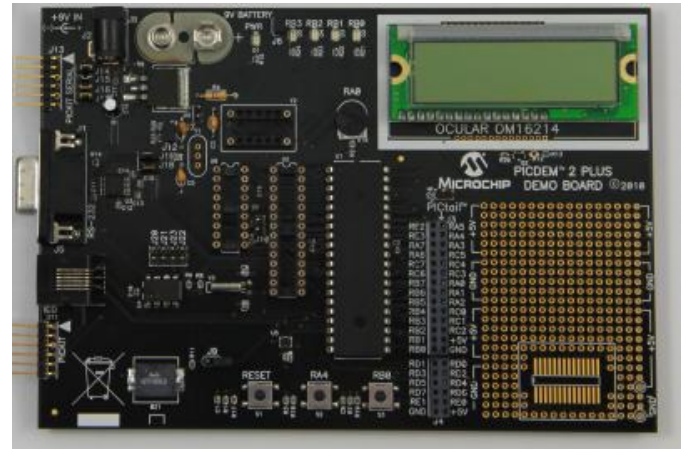

**Fig 2: PIC16F877A Development Kit**

The switching circuit used to ON and OFF the appliances which consist of optical triggered MOC3021 and connected with the PIC16F877A pin. This is used to isolate the load form the controller circuit to prevent any damage cause by the Load. TRIAC IC BTA 116 is used to drive the load which is triggered by the gate current.

Fig.3 shows the proposed algorithm for hardware. At start the hardware initialize and the microcontroller checks for the availability of the Bluetooth device, after this it communicate with the mobile which is the master device with app enabled which receive the voice commands after this it transfer these command to the microcontroller via Bluetooth media which is the heart of this work. After the successful transmission of the commands the controller performs the given operation. And checks for the proper operation of command, on successful operation it will acknowledge the master device that the operation has performed.

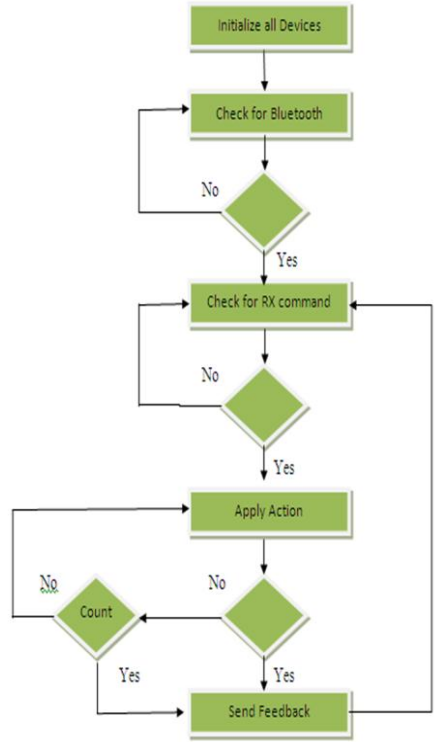

**Fig 3: Flow chart of Proposed Hardware System**

## 4. RESULTS AND DISCUSSION
The proposed system converts the voice commands in to text and sends to the Home mobile via standard GSM SMS architecture; in the given system we test a light, fan and a main switch or main supply as shown in figure 4.

**Table 2: Voice commands and SMS commands with Acknowledgement**

| Voice Command to Text Command Attributes | | | |
|---|---|---|---|
| Voice Commands | SMS Commands | SMS commands Acknowledgement | |
| | | Success | Failure |
| Main Switch On | ' SON1E' | ' SUPPLY 1 on' | ' SUPPLY 1 on 0' |
| Main Switch OFF | ' SOFF1E ' | ' SUPPLY 1 off' | ' SUPPLY 1 off 0' |
| Light On | ' LON1E ' | 'LIGHT 1 on' | 'LIGHT 1 on 0' |
| Light Off | ' LOFF1E ' | 'LIGHT 1 off' | 'LIGHT 1 off 0' |
| Fan On | ' FON1E ' | ' FAN 1 on' | ' FAN 1 on 0' |
| Fan Off | ' FOFF1E ' | ' FAN 1 off' | ' FAN 1 off 0' |

When the user speaks Light on the android OS based Mobile the given application convert the given text in to 'LON1E' format and append the given text in to SMS payload, table 2 shows text against each voice command. When SMS received at the user end and read by the micro controller via Bluetooth module. After the complete reception the concern operation is performed after this the micro controller save the status of the device and sends a feedback on the successfulness or the failure of the concerned operation. Similarly a user can switch off the Light by speaking Light off and the android OS based Mobile application convert the voice in to the 'LOFF1E' text and append it to the SMS payload and the similar process repeats.

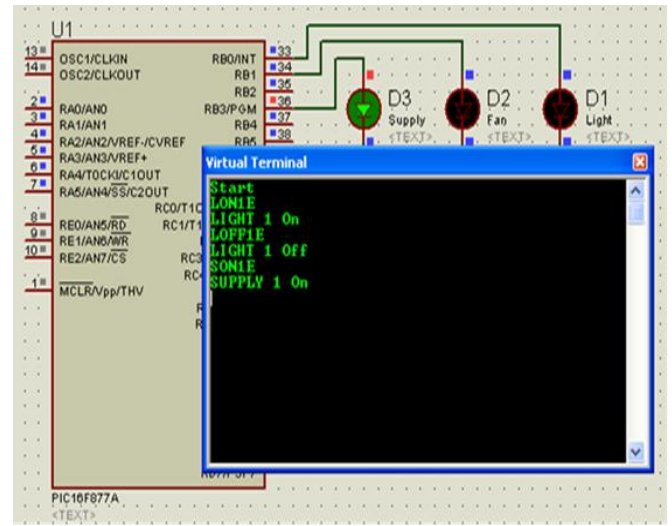

**Fig 4: Voice Command tests for supply and light**

## 5. CONCLUSION
The communication link between the appliances and remote user plays an impotent roll in automation. In this study we proposed a system that control electric appliance via voice when the user is in remote area, and also it controls the appliances through home mobile.